\documentclass[11pt,twoside]{article}

\usepackage{asp2006}
\usepackage{psfig}
\usepackage{epsf}
\usepackage{graphics}
\usepackage{lscape}
\usepackage{array}
\usepackage{amsmath}
\usepackage{amsfonts}

\begin{document}
\title{Realistic Magnetohydrodynamical Simulation of Solar Local Supergranulation}
\author{Sergey D. Ustyugov}
\affil{Keldysh Institute of Applied Mathematics, 4, Miusskaya sq., 
Moscow,
Russia}

\begin{abstract}

Three-dimensional numerical simulations of solar surface magnetoconvection using realistic model
physics are conducted. The thermal structure of convective motions into the upper radiative 
layers of the photosphere, the main scales of convective cells and the penetration depths of convection are investigated. We take part of the solar photosphere with size of $60\times60$ Mm in horizontal 
direction and by depth 20 Mm from level of the visible solar surface. We use a realistic initial 
model of the Sun and apply equation of state and opacities of stellar matter. 
The equations of fully compressible radiation magnetohydrodynamics with dynamical viscosity 
and gravity are solved. We apply:
1) conservative TVD difference scheme for the magnetohydrodynamics,
2) the diffusion approximation for the radiative transfer,
3) dynamical viscosity from subgrid scale modeling.
In simulation we take uniform two-dimesional grid in gorizontal plane and nonuniform grid in vertical 
direction with number of cells $600\times600\times204$. We use 512 processors with 
distributed memory multiprocessors on supercomputer MVS-100k in the Joint Computational Centre of the 
Russian Academy of Sciences.

\end{abstract}
\thispagestyle{plain}

\section{Introduction}

The convection near solar surface develops on the different space and time scales.
Numerical simulation provides useful information about evolution of convective 
structures and helps to construct consistent models of the physical processes    
underlying of observed solar phenomena. We conduct an investigation of the 
temporal evolution and growth of convective modes on scales supergranulation in a 
three-dimensional computational box with imposed initially uniform weak magnetic field. 
In previous work by the author [\citet{ustyugs06}] it was shown that collective 
motions of small convective cells of granulation expels weak magnetic field on the 
edges cells at mesogranular scales. Average size of such cells is 15-20 Mm and the 
lifetime is 8-10 solar hours.  In article [\citet{rast06}] 
it was suggested that development of large spatial and long temporal mesogranular 
and supergranular scales naturally arise as result the collective advective interaction of many 
small-scale and short-lived granular plumes. Simulation of solar photosphere convection 
[\citet{stein06}] in a computational domain of size 48 Mm in horizontal plane 
and 20 Mm in depth showed that the sizes of convective cells increase with depth. 
The purpose of this work is to investigate the effect magnetic field on the development 
of convection on scale supergranulation in a region of size 60 Mm in the horizontal plane 
and 20 Mm in depth. 

\section{Numerical method}

As initial state we take the distribution of the main thermodynamic variables 
with radius from the Standard Solar Model [\citet{christ03}] 
with parameters $(X,Z,\alpha)=(0.7385,0.0181,2.02)$, where $X$ and $Y$ are hydrogen
and helium abundances by mass, and $\alpha$ is the ratio of mixing
length to pressure scale height in the convection region. We used the OPAL opacity 
tables and the equation of state for solar matter [\citet{rogers96}].

The compressible nonideal magnetohydrodynamics equations are solved:




\begin{equation}
\frac{\partial \rho}{\partial t}+ \nabla \cdot \rho \vec v = 0
\end{equation}

\begin{equation}
\frac{\partial \rho \vec v}{\partial t}+ \nabla \cdot \left[\rho \vec v \vec v
+\left( P +  \frac {B^2}{8 \pi}\right)I - \frac {\vec B \cdot \vec B}{4 \pi}\right]
= \rho \vec g + \nabla \cdot \hat \tau
\end{equation}

\begin{eqnarray}
\frac{\partial E}{\partial t}+ \nabla \cdot \left[\vec v
\left(E + P +  \frac {B^2}{8 \pi}\right) - \frac {\vec B \left(\vec v \cdot \vec B\right)}
{4 \pi}\right]  = \frac {1}{4 \pi}\nabla \cdot \left(\vec B \times \eta \nabla 
\times \vec B\right)\nonumber \\ +\nabla \cdot \left(\vec v \cdot \hat \tau \right) 
+ \rho \left(\vec g \cdot \vec v \right) + Q_{rad}
\end{eqnarray}

\begin{equation}
\frac{\partial \vec B}{\partial t}+ \nabla \cdot \left(\vec v \vec B - \vec B \vec v\right)=
- \nabla \times \left(\eta \nabla \times \vec B \right)
\end{equation}

here $\rho$ is the density, $P$ is the pressure, $\vec v$ is vector of the velocity, $\vec B$ is vector 
of the magnetic field, $\vec g$ is vector of the gravitation, $E=e+\rho v^2/2+B^2/8 \pi $ is 
the total energy, $Q_{rad}$ is the energy transferred by radiation and $\tau$ is the viscous stress tensor. 
The influence of small scales on large scale flows was evaluated in terms of
viscous stress tensor and the rate of dissipation was defined from buoyancy and shear 
production terms [\citet{canuto94}]. The evolution of all variables in time 
was found using an explicit TVD[Total Variation Diminishing] conservative difference scheme
([\citet{yee90}])

\begin{equation}
U^{n+1}_{i,j,k}=U^n_{i,j,k}-\Delta tL(U^n_{i,j,k}),
\end{equation}

where $\Delta t=t^{n+1}-t^n$ and the operator $L$ is

\begin{eqnarray}
L(U_{i,j,k})=\frac {\tilde F_{i+1/2,j,k}-\tilde F_{i-1/2,j,k}}{\Delta 
x_i}
+ \frac {\tilde G_{i,j+1/2,k}-\tilde G_{i,j-1/2,k}}{\Delta y_j}\nonumber \\
+ \frac {\tilde H_{i,j,k+1/2}-\tilde H_{i,j,k-1/2}}{\Delta 
z_k}+S_{i,j,k} \ .
\end{eqnarray}


The flux along each direction was defined by the local-characteristic method as 

\begin{equation}
\tilde F_{i+1/2,j,k}=\frac {1}{2}\left[F_{i,j,k}+F_{i+1,j,k}+
R_{i+1/2}W_{i+1/2}\right] \ .
\end{equation}

$R_{i+1/2}$ is the matrix whose columns are right eigenvectors of
$\partial F/\partial U$ evaluated as a generalized Roe average 
of $U_{i,j,k}$ and $U_{i+1,j,k}$ for real gases.
$W_{i+1/2}$ is the matrix of numerical dissipation. The term $S_{i,j,k}$ represents 
the effect of gravitational forces and radiation. Time step integration is by 
third-order Runge-Kutta method [\citet{shu88}]. This scheme is second-order in space 
and time. Central differences were used for the viscous term, and the diffusion approximation
was applied for the radiative term in the energy equation. We used a uniform grid 
in the $x$ and $y$ directions and a nonuniform grid in the vertical $(z)$ direction. 
Periodic boundary conditions were used in the horizontal directions and 
the top and bottom boundary conditions were choosen to be

$$v_{z,k}=-v_{z,k-1}, v_{x,k}=v_{x,k-1}, v_{y,k}=v_{y,k-1}$$
$$dp/dz=\rho g_{z}, p=p(\rho), e=const$$
$$B_{x}=B_{y}=0, dB_{z}/dz=0$$

that is, reflection for the $z$ component of velocity, outflow for the $x$ and $y$ components. 
The pressure and density were derived from the solution of the hydrostatic
equation, using the equation of state by constant value of the internal energy.

\section{Results}

We imposed initially the homogeneuos vertical magnetic field with strength of 50 G. 
For the magnetic diffusivity we take constant value 
$\eta = 1.1\times10^{11}$cm${}^2$sec${}^{-1}$. We carried out calculations in 
the region of $60\times60\times20$ Mm on the grid with $600\times600\times204$ 
cells during of 24 solar hours. On the figures 1-3 the results of the MHD numerical
simulation of development of the convection in the horizontal plane near solar 
surface are presented. We find that the magnetic field concentrates in extended regions 
similar sunspot with diameter about 5 Mm and in thin sheets on the boundary of supergranular 
cells. In these regions we reveal the increase of strength of the magnetic field to values
700-800 G, the decrease of temperature on few thousand kelvin, the small values of the vertical 
component of velocity in average about 0.05 km/sec and as result full absence of the 
convective motion. The magnetic pressure prevents the inflow of the material from outside of 
sunspot and thus the transfer of radiation energy is suppressed. Inside of supergranular cell 
the value of vertical component of the magnetic field is very small about 1 G and less. 
Diverging convective flows from the center supergranular expels weak magnetic field 
to the edges of convective cell. An average size of the supergranular cells is 20-30~Mm.
A material moves from the center of supergranular with the velocity of about 1-1.5 km/sec. 
The maximum value of the magnetic field in the computational domain is 2000 G. 

Inside of supergranular we have a typical picture of evolution of convection
on the scale of granulation with average sizes of the cells about 1.5 Mm and with
lifetime about 4-5 minutes [\citet{nordlund00}]. Here we see wider upflows of warm, 
low density, and entropy neutral matter and downflows of cold, converging 
into filamentary structures, dense material. We observe a continuous picture of
formation and destruction of granules. The granules with highest pressure 
grow and push matter against neighboring granules, that then shrink 
and disappear. Ascending flow increases the pressure in the center of granule 
and upflowing fluids decelerates motion. This process reduces heat 
transport to surface and allows the material above the granule to cool, 
become denser, and by action of gravity to move down. We observe a formation of
new cold intergranule lane splitting the original granule.

From figure 4 we can see properties of turbulent convection on depth 0-4 Mm. 
In this region the cold blobs of the matter moves down by maximum of velocity 
of about 4 km/sec and the Mach number M=0.6. The downdrafts has different and 
very complicate vertical structure. One part of the downdrafts travels on small distance 
from the surface and becomes weak enough to be broken up by the surrounding 
fluid motion. Another part of the downdrafts conserves a motion with high velocity and 
moves on the distance about 6 Mm. We observe that different nature of such 
behavior is provided by the initial conditions of downdrafts formation. 
There are places on solar surface where the material moves from different sides to one 
point and here we find powerful vorticity motions of matter. Due to action
of strong compressibility effects we observe quick output of internal energy
and formation of downdrafts with maximum of values of the vertical velocity.
We detect extended regions with size about 5 Mm in a diameter where turbulent
convection is fully disappeared. 
 
On the depthes from 5~Mm to 8~Mm  we reveal more quiet character of the convective 
flow than in the turbulent zone. Below 8~Mm we see clearly the separate large 
scale density fluctuations and the streaming flow of the material which similar 
to jets with the largest value of the average velocity equal to 1 km/sec. 
In these places the magnetic field has value about 300 G. Besides we find on 
depthes bigger than 10 Mm the separate regions of localization of the huge magnetic 
energy with values of magnetic field in range from 800 to 1000 G (Fig.5). Distances 
between these places with strong concentration of the magnetic field is comparable with 
scales development of the local solar supergranulation. The convective motion in different 
parts of photosphere leads to continuous changes of topology of the magnetic field. 
We find inside of the supergranular cell many magnetic loops with big negative values of 
the vertical component of the magnetic field. 

I am grateful Mausumi Dikpati and Local Organize Committee  for
financial support for my participation in the conference GONG 2008/SOHO XXI.

\begin{figure}[!ht]
  \plotfiddle{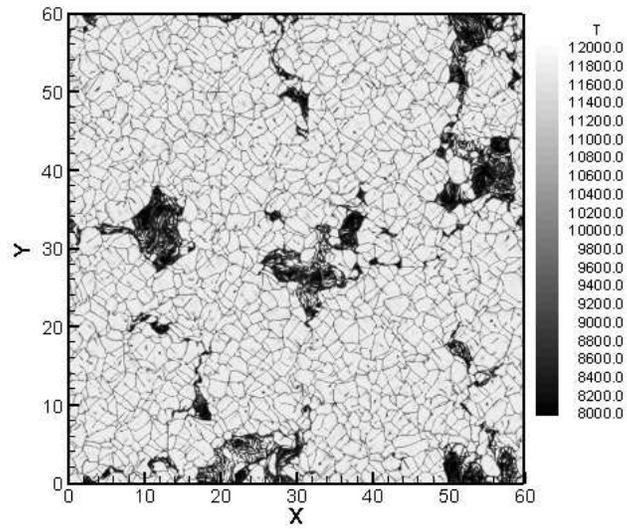}{40mm}{-90}{40}{40}{-150}{180}
  \vspace{2.0cm}
  \caption{\small The contours of temperature on a horizontal plane
near solar surface. The units of temperature is Kelvin. }
\end{figure}

\begin{figure}[!ht]
  \plotfiddle{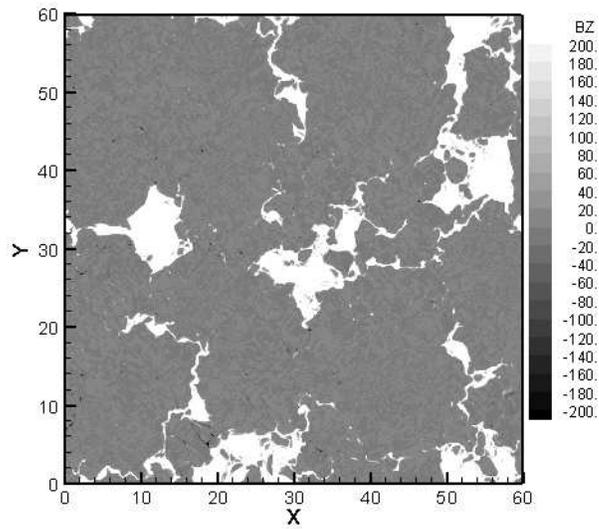}{40mm}{-90}{40}{40}{-150}{180}
  \vspace{2.0cm}
  \caption{\small The contours of vertical component of magnetic field 
on a horizontal plane near solar surface. The units of magnetic field is Gauss. }
\end{figure}

\begin{figure}[!ht]
  \plotfiddle{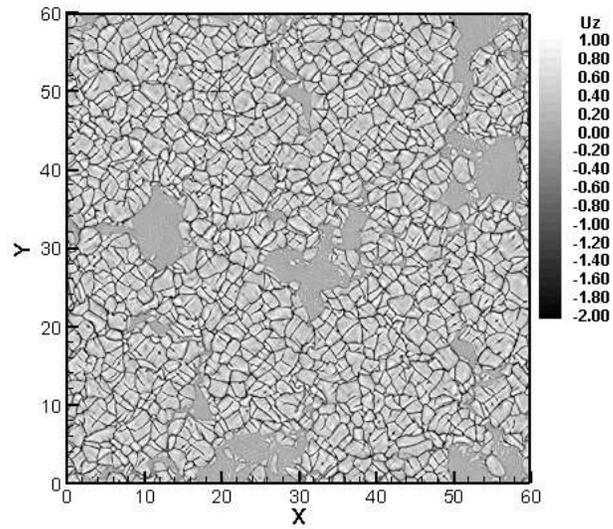}{40mm}{-90}{40}{40}{-150}{150}
  \vspace{2.5cm}
  \caption{\small The contours of vertical component of velocity 
on a horizontal plane near solar surface. The units of velocity is km/sec. }
\end{figure}

\begin{figure}[!ht]
  \plotfiddle{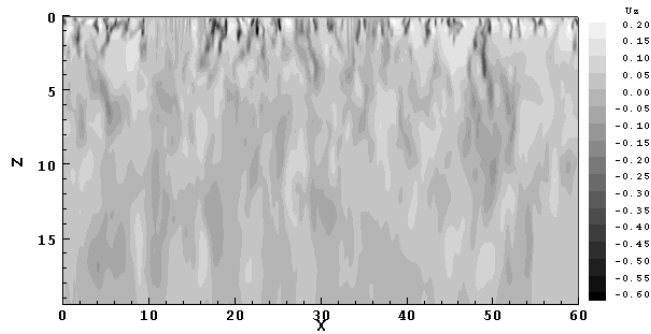}{30mm}{-90}{40}{40}{-150}{160}
  \vspace{1.5cm}
  \caption{\small The contours of vertical component of velocity 
in a vertical plane. The units of velocity in value of speed of sound. }
\end{figure}

\begin{figure}[!ht]
  \plotfiddle{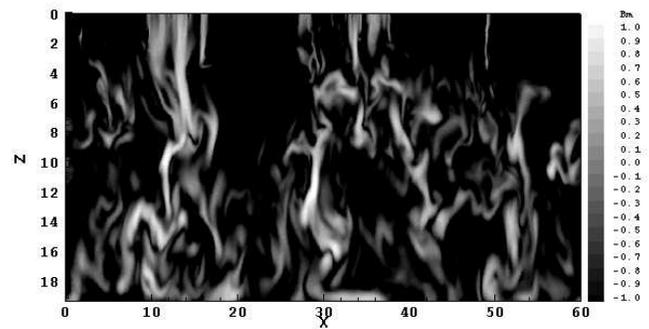}{30mm}{-90}{40}{40}{-150}{160}
  \vspace{1.5cm}
  \caption{\small The contours of magnetic energy (in nondimensional units) in a vertical plane.}
\end{figure}

\end{document}